\documentclass[useAMS,usenatbib]{mn2e}

\usepackage{enumerate}
\usepackage{multirow}
\usepackage{graphicx}
\usepackage{natbib}
\usepackage{longtable}
\usepackage{rotating}
\usepackage[usenames,dvipsnames]{color}

\title[stochastic heating]{Effect of stochastic grain heating on cold dense clouds chemistry}
\author[Chen et al.]{Long-Fei Chen$^{1,2}$, Qiang Chang$^{1}$, Hong-Wei Xi$^{2,3}$\\
$^{1}$Xinjiang Astronomical Observatory, Chinese Academy of Sciences, 150 Science 1-Street, Urumqi 830011, China\\
$^{2}$University of Chinese Academy of Sciences, Beijing 100049, China\\
$^{3}$National Astronomical Observatories, Chinese Academy of Sciences, Beijing 100101, China}

\begin{document}


\maketitle

\label{firstpage}

\begin{abstract}
The temperatures of dust grains play important roles in the chemical evolution of molecular clouds.
Unlike large grains, the temperature fluctuations of small grains induced by photons may be significant.
Therefore, if the grain size distribution
is included in astrochemical models, the temperatures of small dust grains may not be assumed to be constant.
We simulate a full gas-grain reaction network with a set of dust grain radii using the classical 
MRN grain size distribution and include the temperature fluctuations of small dust grains. 
Monte Carlo method is used to simulate the real-time
dust grain's temperature fluctuations which is caused by the external low energy photons
and the internal cosmic ray induced secondary photons.
The increase of dust grains radii as ice mantles accumulate on grain surfaces is also included in our models.
We found that surface CO$_2$ abundances in models with grain size distribution and temperature fluctuations
are more than one order of magnitude larger than those with single grain size. 
Small amounts of terrestrial complex organic molecules (COMs) can also
be formed on small grains due to the temperature spikes induced by external low energy photons.
However, cosmic ray induced secondary photons
overheat small grains so that surface CO sublime and less radicals 
are formed on grains surfaces, thus the production of surface
CO$_2$ and COMs decreases by about one order of magnitude. 
The overheating of small grains can be offset by grain growth so that the formation of surface CO$_2$ and COMs
becomes more efficient.
\end{abstract}

\begin{keywords}
ISM: abundances -- ISM: molecules
\end{keywords}


\section{Introduction}
It has long been recognized that interstellar dust grains play important roles
for the formation of various molecules in various astronomical sources~\citep{Gould1963,Charnley1992,Garrod2006}.
For instance, the efficient conversion from atomic hydrogen to molecular hydrogen 
was believed due to the interaction of
atomic hydrogen and dust grains~\citep{Gould1963,Hollenbach1971,Biham2001}. A well accepted H$_2$ formation 
mechanism is that hydrogen atoms first absorb on the grain surfaces, and then recombine
to form H$_2$ by the Langmuir-Hinshelwood mechanism.
Many other species, including complex organic molecules (COMs), are believed be formed by the same mechanism
on grain surfaces~\citep{Garrod2006,Herbst2009}.

Although dust grains have a size distribution in real astronomical sources \citep{Mathis1977, Draine2001},
most astrochemical models adopt a constant dust grain radius, which is $0.1 ~\mu m$, for the purpose of simplicity.
\citet{Pauly2016}  explained in detail why this approximation generally works reasonably well, but they also argued that
there are limitations introduced by this approximation.
Recently, the grain size distribution has been introduced into astrochemical modeling~\citep{Acharyya2011,Pauly2016,Ge2016}. 
It was found that surface chemistry on grains with a size distribution differ
from that on grains with single radius $0.1~ \mu m$.
On the one hand, with the deposition of the gas phase species, the ice mantle on dust grains grows,
thus the grain radius increase with time, which will increase the effective granular surface area and
the rate of depletion of molecules \citep{Acharyya2011}.
On the other hand, smaller grains will have more surface area because of their larger population.
\citet{Pauly2016} showed that this effect can lead to the majority of ice mantle species
reside on the smallest grains in the core collapse phase.

Perhaps the most significant consequence of dust grain size distribution is the temperature difference of dust grains.
Because the rate coefficients of surface reactions are strongly dependent on the dust grain temperatures,
the ice mantle compositions are dependent on grain sizes. For instance,
smaller dust grains are hotter than larger grains, therefore, species such as CO$_2$ are more likely formed on smaller
grains~\citep{Pauly2016}. 
Moreover, smaller grains have more significant temperature
fluctuations, which is called stochastic grain heating or single photon heating \citep{Draine2001}.
Due to the small size, small grains have a smaller time-averaged vibrational energy \citep{Draine2001}, 
so that grains whose radii are as small as $0.005~ \mu m$ can be heated up to a few tens of kelvins by starlight photons,
thus, the grain temperatures fluctuate between $5 $ K and $50 $ K at $A_v=0~mag$~\citep{Cuppen2006}.
To the best of our knowledge, so far astrochemical models that include stochastic heating of dust grains 
are only limited to the simplest
molecular hydrogen formation models~\citep{Cuppen2006,Bron2014}.
Stochastic heating may give us more new insights into the chemical evolution of ice mantle on dust grains.
The temperature of the hottest grains in quite dark clouds in the models by \citet{Pauly2016} is only
15.6 K while the transient temperature spikes induced by starlight can be more than 20 K.
One particular interesting question is whether COMs, which were
recently detected in cold prestellar cores~\citep{Oberg2010,Bacmann2012}, can be formed on dust grains in cold cores because
of the temperature spikes. The radicals that recombine to form COMs will not be mobile on grain surfaces
until the temperatures of dust grains are more than 20 K~\citep{Garrod2006}, therefore, transient temperature spikes
that are more than 20 K may help to form COMs in cold cores.

In this paper, for the first time, we simulate a full gas-grain chemical network with physical conditions pertain to
cold dark cloud and include more explicit grain temperature fluctuations in our astrochemical models. 
The paper is organized as the following.
The radiation inside cold cores will be introduced in Section \ref{sec:radiation}
while the dust grain-size distribution and grain growth will be explained in Section \ref{sec:distribution}.
Heating and cooling of the grain is introduced in Section \ref{sec:heating}.
The chemical models and simulation method will be introduced in Section \ref{sec:method}.
We will show our results in Section \ref{sec:res}.
The comparison with observations and other models will be discussed in Section \ref{sec:compare}.
Finally, in Section \ref{sec:sum}, we summarize our work and highlight the main conclusions.

\section{Photon Flux in Cold Cores}\label{sec:radiation}

Following \citet{Cuppen2006}, there are two types of photons based their wavelength. Low energy photons are those
whose wavelength are between 250 nm and 1 cm while high energy photons are those whose wavelength are 91.2-250 nm.
Previous studies show that the average temperatures of dust grains are mainly controlled by external radiation field~\citep{Evans2001} and
the heating of dust grains by internal radiation field was ignored~\citep{Zucconi2001}.
However, for a cold dark molecular cloud with extinction $A_v=10~ mag$, external high energy photons can hardly penetrate
into the inner part of the cloud, so the cosmic ray induced secondary photons are the dominant 
high energy photons inside cold cores~\citep{Gredel1989}.
On the other hand, high energy photons carry more energy, so they can cause more significant fluctuations. 
Because surface chemical reactions are very sensitive to dust grain temperature, we
do consider the effect of high energy photons heating in models as explained in the chemical model subsection.

Dust grains are assumed to be silicate,  which is used to choose the absorption coefficient data 
for the dust grains as in \citet{Cuppen2006}.
We follow \citet{Cuppen2006} to calculate the low energy photon flux.
The high energy photon flux in our  models are explained in detail here.
We only consider the cosmic ray induced secondary photons
for high energy photons because they are the dominant ones.
The spectrum of the cosmic ray induced secondary photons is complicated and
consists of many lines~\citep{Shen2004}. So, \citet{Shen2004} smoothed the spectrum between 850 \AA~to 1750 \AA~for
the purpose of simplicity. Moreover, high energy photons can do more than heating dust grains. High energy
photons can photodissociate ice mantle species or desorb surface species into gas phase
so that the energy carried by high energy photons is stored as chemical energy. Assuming one high energy photon
can only either photodissociate or desorb one surface species~\citep{Chang2014}, the rate of high energy photon flux
that is stored as chemical energy is $\sum{rate_i} + \sum{rate_j}$, where i is for all photodissociation reactions and j is for all
photo desorption reactions. For a dust grain with radius $r$, the rate of cosmic ray induced photons 
bombardment is $Q_{abs}G_0F_0 \pi r^2 $,
where $G_0 = 10^{-4}$ is the scaling factor for the cosmic ray induced photons~\citep{Shen2004},
$F_0 = 10^8 cm^{-2}s^{-1}$ is the standard interstellar radiation field and $Q_{abs}$ is the wavelength dependent absorption coefficient
for different sizes of silicate dust grains~\citep{Draine1984,Draine1985}, which is available 
online\footnote{http://www.astro.princeton.edu/$\sim$draine/dust/dust.diel.html}.
So, a bare grain with radius $r$ is bombarded by $Q_{abs}G_0F_0 \pi r^2 \Delta t$ high energy photons within the time period $\Delta t$.
The rate of heating a dust grain with radius $r$ by the cosmic ray induced photons
is $R_{pho} = Q_{abs}G_0F_0\pi r^2 - \sum{rate_i} - \sum{rate_j} $.
Because the spectrum of cosmic ray induced photons is complicated and we are only interested in the
ice mantle composition change due to the temperature spikes induced by high energy photons,
we use the median value between 850 \AA~and 1750 \AA, 1300 \AA~as the wavelength of cosmic ray induced photons in our models.
In order to evaluate the impacts by the approximation, we also assume that
the wavelengths of cosmic ray induced photons are uniformly 
distributed between 850 \AA~and 1750 \AA~ in one model.

\section{Grain Size Distribution and Grain Growth}\label{sec:distribution}
Dust grains in dense clouds may coagulate~\citep{Chokshi1993,Ysard2016}. 
So, the distribution of dust grain size in dense
clouds may differ from the classical MRN grain size distribution~\citep{Mathis1977}, 
which is representative of the diffuse interstellar medium. However,
the MRN size distribution has a simple analytical formula.
Moreover, the grain size distribution in dense clouds is much more poorly known than in diffuse clouds. 
So, the MRN grain size distribution was still used by \citet{Pauly2016} and \citet{Acharyya2011} to 
study the effects of grain size distribution on the chemical evolution of dense clouds because  
simulation results based on the MRN size distribution can help as a first step to elucidate these effects.
Therefore, we also assume that the grain-size 
distribution follows the classical MRN size distribution \citep{Mathis1977},
which is $dn \propto r^{-3.5}dr$. The largest and smallest radius are 0.25 $\mu m$ and 0.005 $\mu m$ respectively.
Following \citet{Pauly2016} and \citet{Acharyya2011}, we initially divided the size distribution
into five logarithmically equally spaced bins across the range of cross-section area.
Then we calculate the mean cross-section area for each bin. Finally, the representative 
grain radius for each bin
is calculated  based on the mean cross-section area for that bin~\citep{Pauly2016}.
The 0th bin is the bin with $r_{max0} =0.25 ~\mu m$.
The number of grains simulated for each bin follows the relative abundances of grains according to the MRN dust
distribution. The reason will be discussed in Section~\ref{sec:method}.
We set the number of the grains in the 0th bin to be 1, so the number of dust grains in the ith bin, $N_i$ is calculated as,
\begin{equation}
 N_i = \frac{\int_{r_{mini}}^{r_{maxi}} r^{-3.5}dr}{\int_{r_{min0}}^{r_{max0}} r^{-3.5}dr}
\label{num}
\end{equation}
The initially calculated representative radii for the 0th through 4th bins
are $0.157~\mu m$, $0.0718~\mu m$, $0.0328~\mu m$, $0.0150~\mu m$ and $0.00687~\mu m$ respectively.
The numbers of dust grains in the 0th through 4th bins are found to be 1, 7, 49, 353 and 2500 respectively.
Totally there are almost 3000 dust grains.
The computational cost to simulate the chemical evolution of a chemical system with almost 3000 dust grains
and the associated gas phase species is very expensive because we have limited CPUs.

However, the number of total dust grains can be reduced based on previous studies in order to reduce computational cost.
\citet{Garrod2011} found surface chemistry is robust
as long as the grain temperatures are not above 12 K or below 8 K.
We found that the temperatures of grains in
the 0th, 1st and 2nd bins rarely fall out of the 8-12 K range if these grains
are only heated by external low energy photons. 
On the other hand, taking into account of the internal high energy photons heating,
the temperatures of grains in the 2nd bin can exceed 12 K while the temperatures of grains in the 0th and 1st bins
are still within the range 8-12 K. So the temperature fluctuations of the dust grains in the 0th, 1st and 2nd bins
can be ignored if we only consider the ambient radiation heating, however, the temperature fluctuations of
the grains in the 2nd bins are significant if the heating by cosmic ray induced photons are considered.
The surface chemistry on smaller dust grains may be different than that on much larger dust grains
because the  numbers of reactive species may be much less than one on the smaller grains, thus, the
fluctuations of surface species numbers on smaller grains may change surface chemistry~\citep{Biham2001}. 
However, it was found that the finite size effect is not important for a simple molecular hydrogen 
formation reaction network on dust grains with radius more than 0.05 $\mu m$~\citep{Biham2001} and becomes less important
as the reaction network becomes more complicated~\citep{Chang2014}.
So, we can conclude that surface chemistry on the grains in the 0th and 1st bins are almost identical.
Therefore, we can merge the 0th and 1st bins into a new 0th bin for a valid approximation.
The mean cross-section area for the new 0th bin is calculated in order to calculate the representative 
grain radius for this bin\citep{Pauly2016}.
The number of grains in the new 0th bin is reset to be 1 and the numbers of grains in other bins are calculated
by Equ.~\ref{num}.
Table \ref{tab:grain} summarizes the radii of the representative grains and the number of grains in each bins. 

Grain growth due to the increase of ice mantle is also included in our models. 
With the gas phase species deposition, the grain radius will gradually become larger.
The temperature fluctuation of dust grains will become less as ice mantle gradually becomes thicker.
The depth of a monolayer of a grain is assumed as $d_{ML} = 1/\sqrt{A_s}$ \citep{Pauly2016}, where A$_s$ is the
site density of the grain. For a site density of $1.5\times10^{15} cm^{-2}$, $d_{ML} = 2.582\times10^{-8} cm$.

\begin{table*}
\caption{Calculated grain-size distribution}
\begin{tabular}{lllll}
\hline
bin                                 &  0                        &   1                &  2                   &   3\\
calculated radius/$\mu m$           &  0.086997                 &   0.03283          &   0.01501            &   0.00687   \\
number                              &  1                        &   7                &   44                 &   310         \\
\hline
\label{tab:grain}
\end{tabular}
\end{table*}

\section[]{Heating and Cooling of the Grain}\label{sec:heating}

Dust grains can also be directly
heated by the bombardment of cosmic rays~\citep{Kalvans2016}. However, because the heating by cosmic rays
is less frequent than that by photons, we focus on the temperature fluctuations induced by photons only in this work.
The grain temperature fluctuations are caused by the absorption of low and high energy photons,
which are treated as discrete random events in our simulations.

The heat capacity is necessary to calculate the temperatures of grains that absorb photons. 
Following \citet{Cuppen2006}, in which all the interstellar dust grains are assumed to be spherical silicate grains, 
the size dependent
heat capacity to convert photon energies to temperature spikes is,
\begin{equation}
c(T) = 61.38 r^3 T^3,
\label{equ0}
\end{equation}
where r is the radius of a dust grain in $\mu$m while T is the temperature of a dust grain.
The heat capacity is in eV K$^{-1}$.

Only radiative cooling of the grain is considered in our models in this work. 
The radiative cooling is treated with the usual continuous cooling approximation.
We follow \citet{Cuppen2006} to calculate the temperatures 
of dust grains by radiative cooling. The sublimation of CO can also take away energy~\citep{Schutte1991}, 
however, CO sublimation cooling is not included in our models.
We discuss the significance of CO sublimation cooling in Section~\ref{sec:sum}.

\section[]{Chemical Models and Methods}\label{sec:method}

\subsection{Chemical Models}
The chemical reaction network used in this work is from \citet{Hincelin2011}, 
which is a full gas-grain reaction network. We slightly modify the reaction network as the following.
First, the accretion of molecular hydrogen is absent in our reaction network in order to reduce the computational
cost. Indeed, recent study shows that the absence of gas phase H$_2$ accretion can only introduce
negligible errors~\citep{Chang2017}. Second, we use the competition mechanism for the hydrogenation 
reactions that convert CO to methanol~\citep{Chang2007,Garrod2011,Chang2012}. The hydrogenation of CO to form methanol
has been studied extensively in laboratory experiments and the reaction barriers for the reactions 
H + CO and H + H$_2$CO have been fitted using microscopic Monte Carlo method, which naturally includes
the reaction-diffusion mechanism~\citep{Fuchs2009}. The reaction barriers for H + CO and H + H$_2$CO 
in ~\citet{Fuchs2009} are used in this work. We use  a two-phase (gas phase and grain surface) 
model in which no distinction is made between the active layers and ice mantles for simplicity.

The density is fixed to be, n$_H$=$2\times10^4$ cm$^{-3}$ while the gas temperature is fixed to be 10 K.
The extinction is  $A_v=10 mag$. The initial abundances used in this work are from \citet{Semenov2010} 
and are shown in Table~\ref{tab:init}.
We keep the total dust to gas mass ratio to be 0.01. In total, there are N$_H$=$7.11\times10^{11}$ H nuclei in all cells of gas.
Each grain is put in a cell of gas. The number of H nuclei in each cell is proportional to the
cross-section area of the grain in the cell. The reason will be be clear in the method section. 
The temperature fluctuation of dust grains with different sizes can be calculated 
with the methods introduced in ~\citet{Cuppen2006}.
To save CPU time,
we neglect the temperature fluctuation of grains in the 0th bin and keep their temperature to be 10 K.
Moreover, the temperature fluctuation of grains in the 1st bin
induced by the low energy photons are also ignored. In order to take into account the low energy photons heating,
the time-averaged temperature of grains in the 1st bin is calculated by a single run of
temporal grain temperatures  within a period of one year, which is found to be 10.31 K.
The temperature fluctuations of grains in the 1st bin induced by high energy photons are included in models.
All models are simulated for a period of $2\times10^5 yrs$ which is
the earliest time when gas phase species by model results agree best with observations in cold cores.

The five models simulated in this work are summarized in Table~\ref{tab:model}.
Model M1 includes the heating induced by low energy photons only
while models M2 and M3 include both high energy photons (cosmic ray induced UV photons) grain heating
and low energy photons (infrared wavelength photons) grain heating.
The spectrum of the high energy photons are approximated by a single energy (1300 \AA) in model M2,
while the wavelength of high energy photons are assumed to be uniformly distributed between 850 \AA~and 1750 \AA~in model M3.
Grain growth is not included in models M1, M2, M3.
The effect of grain growth is studied in model M4 in which dust grains are heated by both the high and low energy photons
and grow in size as gas phase species accrete on grain surfaces.
We also simulate a reference model, M5 in which stochastic heating and the growth of dust grains are absent.
The radius of dust grains used in model M5 is the standard one, $0.1\mu m$.
The dust to gas mass ratio is also kept to be 0.01 in model M5 so that there are $6.16\times10^{11}$ H nuclei
in the single cell of gas.

\begin{table*}
\caption{Initial Abundances}
\begin{tabular}{ll}
\hline
species &   abundance$^{a}$\\
\hline
$H_2$   &   0.5\\
$He$    &   9.00(-2)\\
$C^+$   &   1.20(-4)\\
$O$     &   2.56(-4)\\
$N$     &   7.60(-5)\\
$P^+$   &   2.00(-10)\\
$Na^+$  &   2.00(-9)\\
$Cl^+$  &   1.00(-9)\\
$Si^+$  &   8.00(-9)\\
$S^+$   &   8.00(-8)\\
$Mg^+$  &   7.00(-9)\\
$Fe^+$  &   3.00(-9)\\
\hline
\label{tab:init}
\end{tabular}
\medskip{\protect\\
Notes.\protect\\
$^{a}$abundance with respect to H. $a(b) = a\times10^b$.}
\end{table*}

\begin{table*}
\caption{Chemical Models}
\begin{tabular}{llll}
\hline
Model                 &  Grain-size Distribution & Heating Source                               &   Grain Growth   \\
\hline
M1                    &   Yes                    & Low Energy Photons                           &   No            \\
M2                    &   Yes                    & Low Energy Photons and                       &   No            \\
                      &                          & High Energy Photons (1300 \AA)               &        \\
M3                    &   Yes                    & Low Energy Photons and                       &  No            \\
                      &                          &  High Energy Photons (850 \AA~to 1750 \AA)    &        \\                      
M4                    &   Yes                    & Low Energy Photons and                       &   Yes           \\ 
                      &                          & High Energy Photons (1300 \AA)               &        \\ 
M5                    &   No                     & None                                         &   No            \\
\hline
\label{tab:model}
\end{tabular}
\end{table*}

We define three terms to be used in the results section.   
The total population of a species, X, in all cells is noted as N(X)
while the total population of the species X in all cells containing grains with the same radii $r_i$ is noted as 
N(X)$_{r_i}$, where i is 0, 1, 2, and 3. 
The radii $r_0$, $r_1$, $r_2$ and $r_3$ are 0.086997$\mu m$, 0.03283$\mu m$, 0.01501$\mu m$ and 0.00687$\mu m$ respectively.
If the species X is a surface species,  N(X)$_{r_i}$ is also the total population of the species X
on all dust grains with radii $r_i$.
The total fractional abundance of a species, X,
is defined as N(X)/N$_H$.
The fractional abundance of a species, X, in a type of cells is defined as 
 N(X)$_{r_i}$/(N$_H$)$_{r_i}$, where (N$_H$)$_{r_i}$ is the total population of H nuclei in 
all cells containing grains with the same radii $r_i$.
The type of cell is represented by the size of dust grains in the cell. 
Finally, the fractional abundance of a surface species, X, 
on grains with radii $r_i$ is defined as  N(X)$_{r_i}$/N$_H$.

\subsection{Methods}

The Gillespie algorithm is an ideal method to study stochastic chemical kinetics~\citep{Gillespie1976}.
However, we found that the next reaction method~\citep{Gibson2000, Chang2012}, which is a modified
Gillespie algorithm, is more suitable to study the surface chemistry on stochastically heated grains.
Therefore, the next reaction method is used in this work. The detailed explanation of the next reaction
method can be found in ~\citet{Gibson2000} and \citet{Chang2012}. We only briefly explain the method in the following.

The absolute time when the next reaction occurs is used in the next reaction method. 
For gas phase reactions or surface reactions on dust grains with constant temperatures,
the ith reaction will occur after time interval, $\tau_i = -\ln X/R_i$ 
where $X$ is a random number uniformly distributed within 0 and 1
while $R_i$ is the reaction rate of the ith reaction. 
So the absolute time when the next ith reaction will occur is $t_i = t_{0i} + \tau_i$
where $t_{0i}$ is the current time. If the reaction rate of the ith 
reaction changes from $R_i$ to $R_i^{'}$ at time $t_{0i}^{'}$,
where $t_{0i}<t_{0i}^{'} < t_i$,
the absolute next reaction time for the ith reaction is updated as $t_i^{'} = t_{0i}^{'} + \frac{(t_i - t_{0i}^{'})R_i}{R_i^{'}}$.
The temperatures of dust grains always gradually cool down except when the grains are bombarded by star photons.
Therefore, for the jth surface reaction on stochastically heated grains, the absolute next reaction time
$t_j$ is calculated by the following equation,
\begin{equation}
-\ln Y = \int_{t_{0j}}^{t_{j}} R_j(T_d(t)) dt, 
\label{tau}
\end{equation}
where $Y$ is another random  number uniformly distributed within 0 and 1, $t_{0j}$ is the current time,
$R_j$ is the reaction rate of the jth reaction and $T_d$ are the time dependent temperatures of dust grains that cool down
by radiative cooling.
Similarly, when the value of $R_j$ changes because dust grains are heated by photons  
or other reactions change the abundances of the reactants of the jth reaction at time $t_{0j}^{'}$, the new absolute
next reaction time $t_j^{'}$ is calculated as,
\begin{equation}
-\ln Y = \int_{t_{0j}}^{t_{0j}^{'}} R_j(T_d(t)) dt + \int_{t_{0j}^{'}}^{t_{j}^{'}} R_j^{'}(T_d(t)) dt
\label{tau2}
\end{equation}
where $R_j^{'}$ is the new reaction rate of the jth reaction.
Following~\citet{Cuppen2006}, we calculate the next arrival time of low energy photons.
The next arrival time of cosmic ray induced photons can be calculated in a manner similar to Equ.~\ref{tau},
using the rate of cosmic ray induced photons heating, $R_{pho}$ discussed before.
Thus, the absolute time when each event occurs can be found.
We sort out the minimum absolute time and then execute the event, which is a chemical reaction or photon heating.
If it is a chemical reaction, we change the population of reactants and products of the reaction and then
update all the rates of reactions that the reactants and products of the reaction participate.
If it is a photon heating event, we increase the temperature of dust grains and
update all the rates of surface reactions whose reaction rates are dependent on surface temperatures.
The process is repeated before the gas-grain system reaches the specified final time.

In order to accelerate simulations, parallel computation is used.
We can use one CPU to simulate the chemical evolution of surface ice and gas in one cell.
In total we need almost 400 CPUs.
However, we found that the computational costs to simulate the chemical evolution
of different cells of gas and surface ice are different. For instance,
the CPU time required by cells which contain grains in the 2nd bin
is much more than that by the smallest cells which contain grains in the 3rd bin.
In order to save CPUs, we use the multi-thread programming to simulate multiple cells of gas and ice in one CPU.
One CPU can handle five of the smallest cells or one any other larger cell.  
Thus, the number of CPUs used is reduced.
Totally 108 CPUs are used to perform simulations in this work.

We assume the gas phase species are well mixed, so the number density of the same gas phase species in different cells of gas
should be the same. Thus, the number of a gas phase species in different cells should be proportional to the volume of the cells.
However, as the chemical system gradually evolves, the number density of gas phase species in different cells may not be the same
because the chemical evolution of different cells may be different.
The number of any gas phase species in each cell of gas is proportional to the surface area of the dust grain in each cell
in order to ensure that the accretion of gas phase species does not introduce any difference of 
number density of gas phase species in different cells. However, chemical reactions other than accretion
may introduce differences in the gas phase species number density.
For instance, because of the temperature spikes on the smallest dust grains, 
surface H atoms are more likely to desorb from the smallest grains than the largest grains. Therefore, 
the number density of gas phase H in the cell of gas that contains the smallest grains may be higher than
that in the cell of gas that contains the largest grain. Thus, we must mix gas phase species in different cells so that 
the number densities of each gas phase species in all cells are the same.  
Suppose we mix gas phase species in $N_1$ cells that contain the smallest grains and $N_2$ cells that contain the largest grains. 
The number density of gas phase species i after mixing is,
\begin{equation}
 \bar{n}(i) = \frac{N_1 n(i)_1 +  N_2n(i)_2}{N_1V_1 + N_2V_2} = \frac{\frac{N_1}{N_2} n(i)_1 +  n(i)_2}{\frac{N_1}{N_2}V_1 + V_2} , 
\label{ave_n}
\end{equation}
where $V_1$ and $V_2$ are the volumes of the cells containing the smallest and largest grains respectively
while $n(i)_1 $ and $n(i)_2 $ are the population of gas phase species i in the cells containing the smallest 
and largest grains respectively.
Equ.~\ref{ave_n} shows that the ratio $N_1/N_2$ cannot be arbitrarily chosen in models because $\bar{n}(i)$ is dependent on $N_1/N_2$.
The relative abundances of grains according to the dust size distribution should be used to determine $N_1/N_2$.

In order to keep the gas phase species well mixed,
the Message Passing Interface (MPI) scheme is used to exchange information between CPUs.
The critical information is the number of each species in each cell.
We redistribute the number of gas phase species in cells so that the number density of any species in different
cells remain the same in simulations.
We call the time interval mixing period later.
On the other hand, the more information exchange between CPUs, the more CPU time is consumed for the information exchange by MPI.
Therefore, the mixing period should be as long as possible. Different mixing periods are used in simulations 
to test the convergence of simulation results.

The Taurus High Performance Computing system of Xinjiang Astronomical Observatory is used for 
the simulations in this work. It takes about one week to simulate models M1, M2 and M3 while the simulation of model M4
takes about two weeks.

\section{Results}\label{sec:res}

\subsection{Grain Temperature Fluctuations and Grain Growth}\label{sec:res_tem}
\begin{figure*}
\includegraphics[scale=0.75]{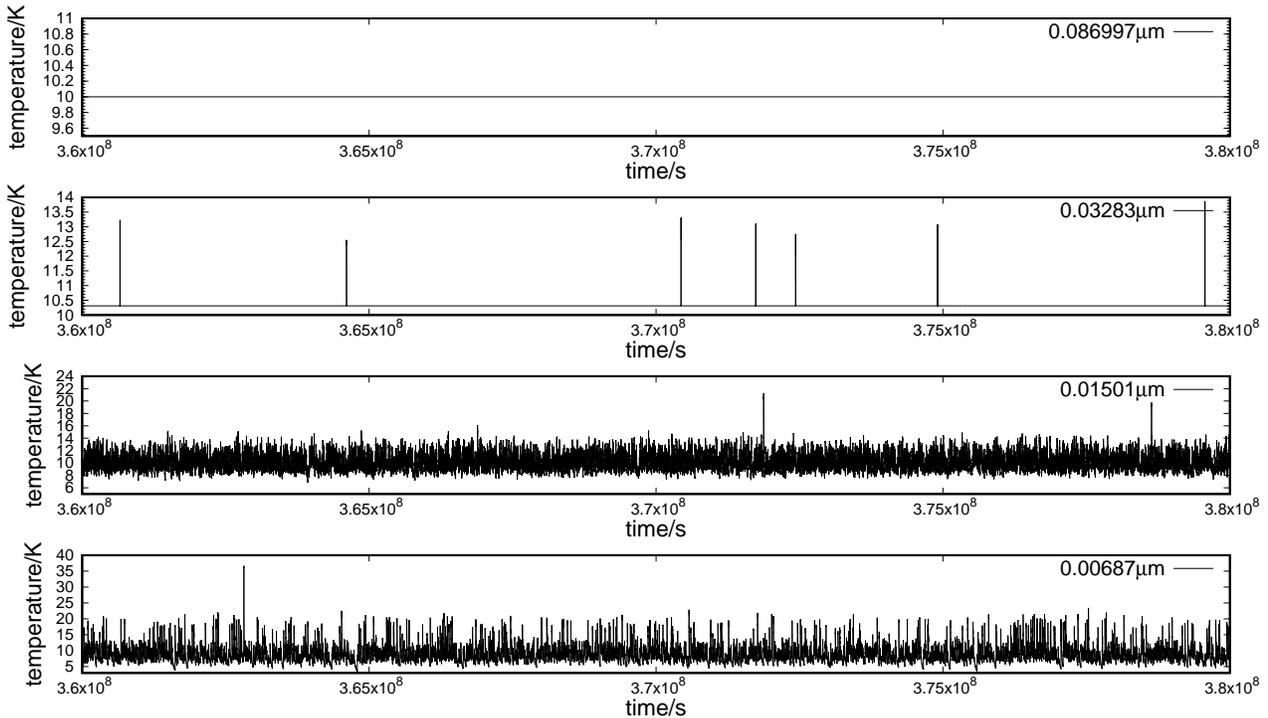}
\caption{The temperatures as a function of time for different sizes of grains in model M3.\protect\\
}
\label{fig:fig_1}
\end{figure*}

Fig.~\ref{fig:fig_1} shows temperature fluctuations for different sizes of grains as a function of 
time in model M3, which includes both high and low energy photons heating. The wavelengths of high energy photons are
uniformly distributed between 850 \AA~and 1750 \AA.
We can see that high energy photons heating events are much rarer than low energy photons heating.
For grains with radius $0.086997\mu m$,
we ignore temperature fluctuations because surface chemistry is robust within the range of 
temperature fluctuations as explained before.
Similarly, for grains with radius $0.03283\mu m$, we only consider the temperature fluctuations induced by
the high energy photons which can heat dust grains up to 13 K.
For grains with radius $0.01501\mu m$ and $0.00687\mu m$, both the high and low energy
photons grain heating are included. 
Low energy photons can heat grains with radii $0.00687\mu m$ and $0.01501\mu m$
to more than 20 K and 14 K respectively. The temperature spikes induced by high energy photons
are much higher than that by low energy photons.  
Grains with radii $0.00687\mu m$ and $0.01501\mu m$ 
can be heated by high energy photons up to temperatures 
more than 35 K and 20 K respectively.

\begin{figure*}
\includegraphics[scale=0.75]{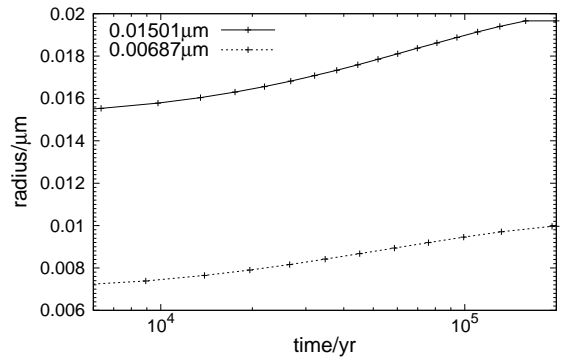}
\caption{The growth of grains with initial radii $0.01501\mu m$ and $0.00687\mu m$ in model M4.\protect\\
}
\label{fig:fig_2}
\end{figure*}

The growth of grain radii is shown in Fig.~\ref{fig:fig_2}, which corresponds to model M4. 
The radii of grains with initial radii $0.01501\mu m$ and $0.00687\mu m$
can increase up to about $0.02\mu m$ and $0.01\mu m$ respectively at the end of simulation.

\begin{figure*}
\includegraphics[scale=0.75]{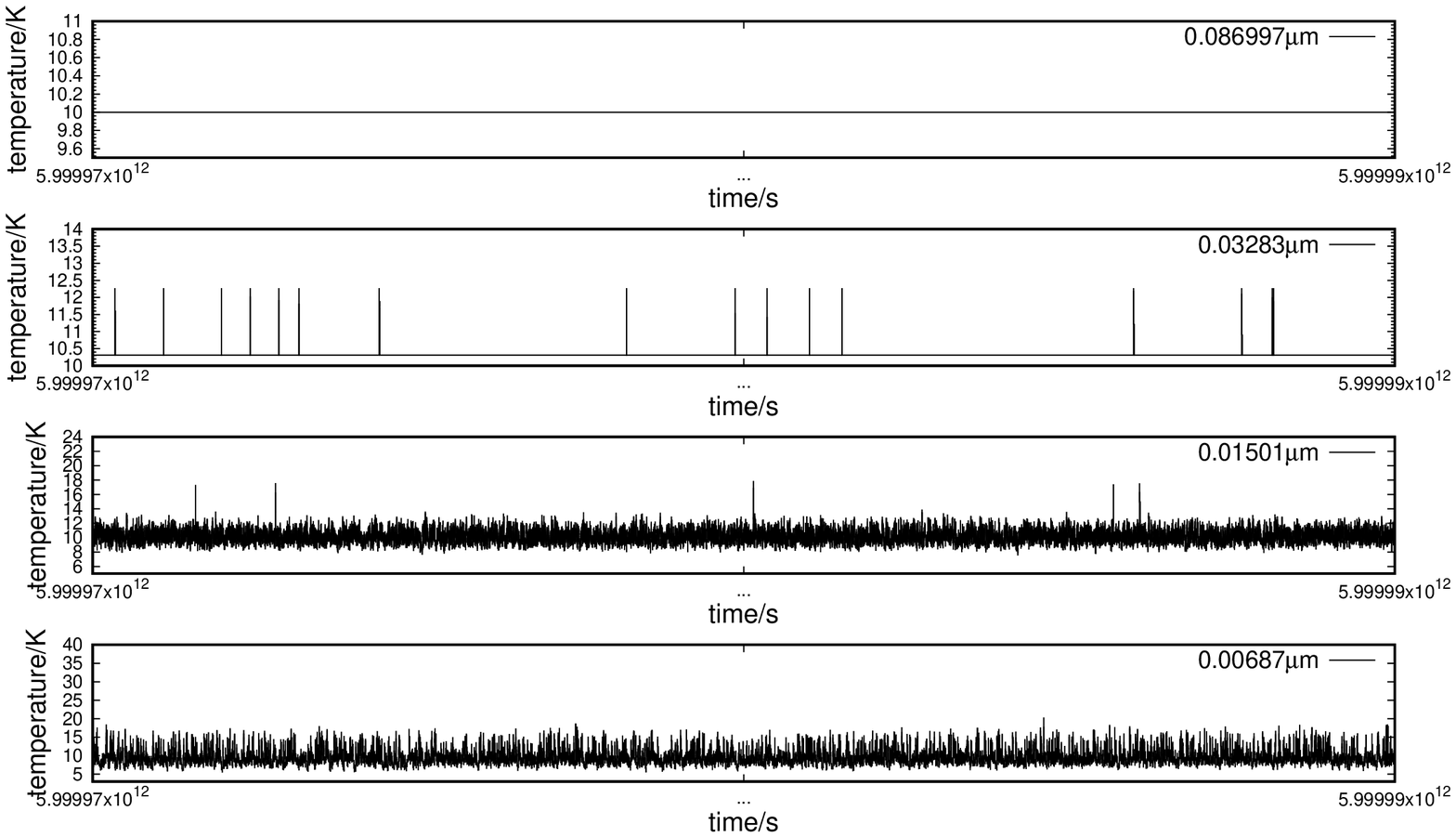}
\caption{The temperatures as a function of time for different sizes of grains in model M4 
at about $1.9\times 10^5$ yrs.\protect\\
}
\label{fig:fig_3}
\end{figure*}

Fig.~\ref{fig:fig_3} shows the temperature fluctuations of dust grains at about $1.9\times 10^5$ yrs in model M4 
which includes the grain growth.
We can see that the maximum temperatures decrease by a few kelvin for grains 
with initial radii $0.01501\mu m$ and $0.00687\mu m$ due to the grain growth.

\subsection{Influence of Mixing Periods }\label{mixing}

First, we test the convergence of simulation results by different mixing periods.
Initially, the number of H$_2$ in a cell is proportional to the volume of the cell.
Moreover, because H$_2$ is overwhelmingly more abundant than any other species,
the formation or destruction of H$_2$ by other species do not change much of the  H$_2$ abundances in each cell.
Thus, H nuclei in each cell is proportional to the volume of the cell. Therefore,
the fractional abundance of a species in a cell is proportional to the number density of the
species. Thus, the fractional abundances in a cell instead of number density 
of gas phase species are used to test the convergence of simulation results.

\begin{figure*}
\includegraphics[scale=0.8]{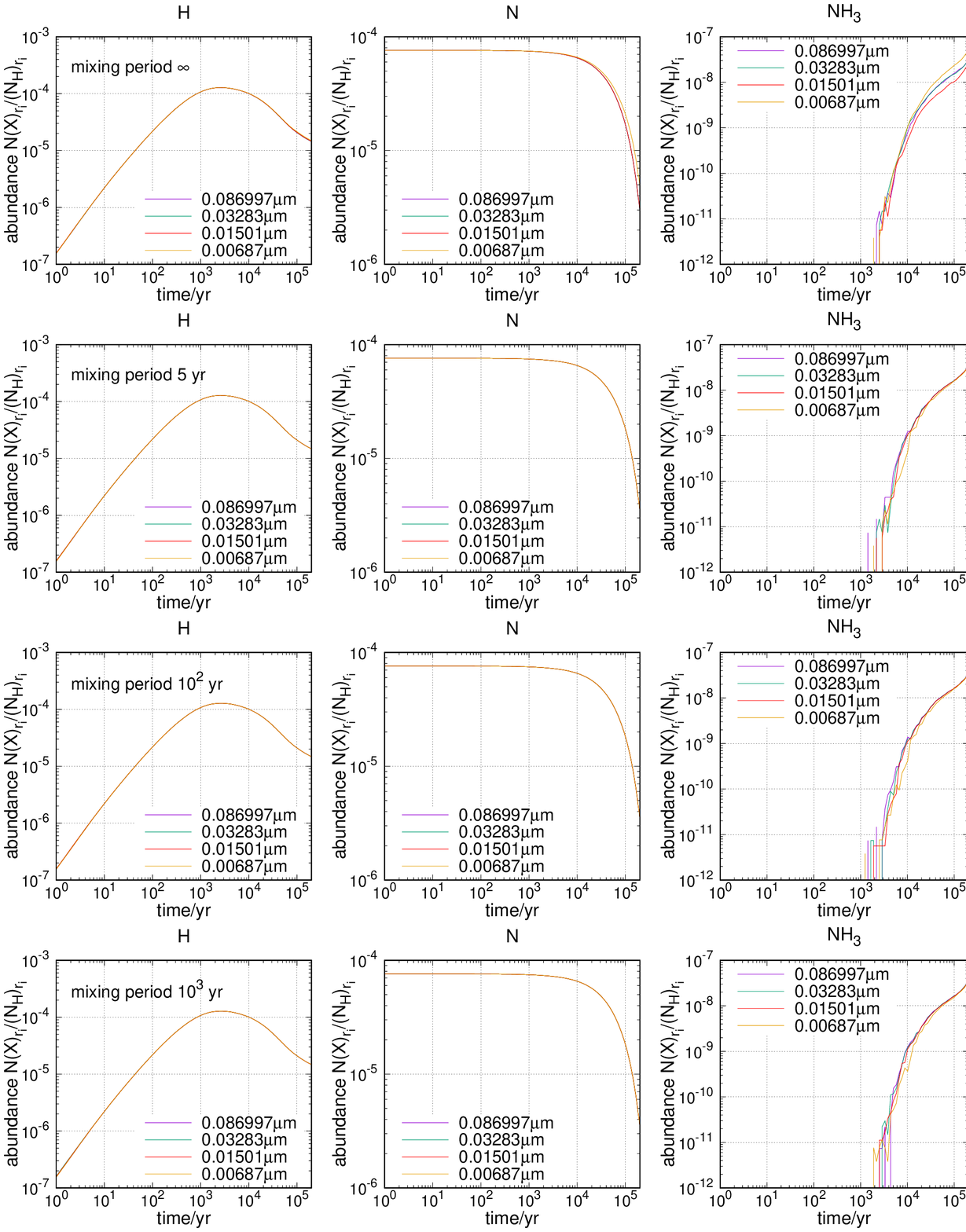}
\caption{Influence of mixing period on the fractional abundance of gas phase species in different cells.
}
\label{fig:fig_4}
\end{figure*}

In order to find out the mixing period that is as long as possible,
model M1 is simulated with mixing periods $10^3$, $10^2$ and $5$ yrs respectively.
We also simulate model M1 with an infinitely large mixing period so that
no information is exchanged between different CPUs.
Fig.~\ref{fig:fig_4} shows the fractional abundances of selected gas phase species
in different types of cells as a function of time for different mixing periods.
The different types of cells are represented by the size of dust grains in the cells.
The influence of different length of mixing period can be seen from Fig.~\ref{fig:fig_4}.
With a mixing period that is infinitely large, because species such as H or N can easily desorb from grain surface
when the temperatures of the smallest grains are well above 10 K,
the fractional abundances of gas phase N in the cells with the smallest grains
is higher than that with the largest dust grains whose temperatures are kept to be 10 K.
For finite values of mixing periods which are no more than $10^3$ yrs,
the fractional abundances of N in different cells
are very close to each other. The fractional abundances of NH$_3$ in different cells are almost identical
as the mixing periods take finite values no more than $10^3$ yrs.
The difference of H abundances in different cells is not obvious even the mixing period is infinitely large.
Our simulation results shows that results do not change much as long as the mixing periods are not more than $10^3$ yrs,
which indicates that $10^3$ yrs is ``short'' enough so that simulation results converge.
So we always use a mixing period of $10^3$ yrs in simulations.

\subsection{Ice Composition and Surface COMs}\label{sec:grain}
\begin{figure*}
\includegraphics[scale=0.8]{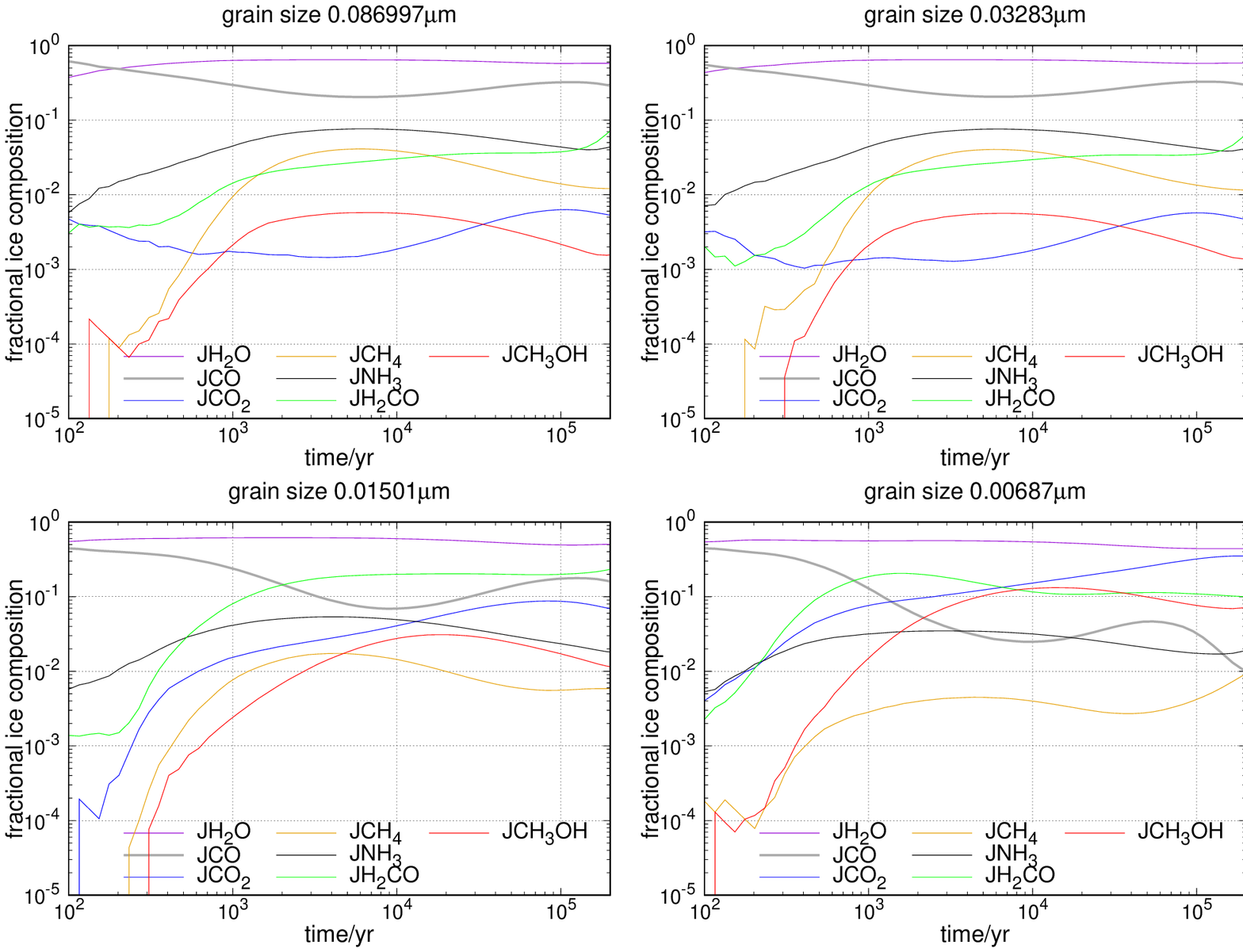}
\caption{The fractional ice composition (showing only major constituents) on different sizes of dust grains as a function of time in model M1.
}
\label{fig:fig_5}
\end{figure*}

Fig.~\ref{fig:fig_5} shows the temporal evolution of
the fractional ice composition (showing only major constituents) for different sizes
of dust grains in model M1. The fractional ice composition is the
percentage of that species with respect to the total number of major species in the ice mantle.  
We use the letter J to designate surface species hereafter.
The major surface species are JH$_2$O, JCO, JCO$_2$, JCH$_4$, JNH$_3$, JH$_2$CO and JCH$_3$OH. 
The fractional ice mantle compositions on the two largest grains are similar in model M1.
Water ice is the most abundant granular species while surface CO is the second most abundant surface species.
There is little JCO$_2$  produced on these two largest grains in model M1 because the temperatures of these grains
are constant and less than 12 K so that the most efficient reaction to form JCO$_2$, JCO + JOH $\rightarrow$ JCO$_2$ + JH, 
does not happen easily. 
As grains become smaller, temperature fluctuations induced by photons are more significant, thus,
the temperature of the smallest and the second smallest may exceed 12 K.
We can see that the fractional ice composition changes significantly. 
First, the fractions of JCO$_2$ on the two smallest grains increase,
and even exceed the fraction of JCO at the end of the evolution on the smallest grains.
Secondly, increasing dust grain temperature can increase the mobility of JH, thus, 
increases the rates at which
JH atoms encounter other species. However, the reactions JH + JCO and JH + JH$_2$CO become more difficult to happen
as dust temperatures increase because we use competition mechanism for these two reactions. 
We can see that the fraction of JCH$_3$OH increase as dust grains become smaller.
The  fraction of JCO decreases on the two smallest grains because JCO molecules are easier to be converted to other
species due to the temperature spikes. 
We found that few JCO molecules can sublime regardless of the size of the grain in model M1.
Finally, 
the fractional ice compositions of JH$_2$O, JCH$_4$ and JNH$_3$ do not change much
as grains become smaller.

The temporal evolution of the fractional ice composition for different sizes of grains reported above
is one realization because we run model M1 only once. 
Different realizations of the time evolution of the fractional ice composition in the same model
might be different. Moreover, the fluctuation of the fractional ice composition in model M2 may be larger
than that in model M1 because of the larger temperature fluctuations in model M2.  
In order to check the convergence of simulation results, we run model M2 five times with different random seeds.
We found that other than species with low abundances at early time, the fractional ice composition for different 
sizes of grains remain almost the same in the five different realizations.
Therefore, the convergence of simulation results is achieved if we simulate each model just once.   
So, model results reported in this work are all from one realization of the temporal evolution of species abundances in each model. 

\begin{figure*}
\includegraphics[scale=0.8]{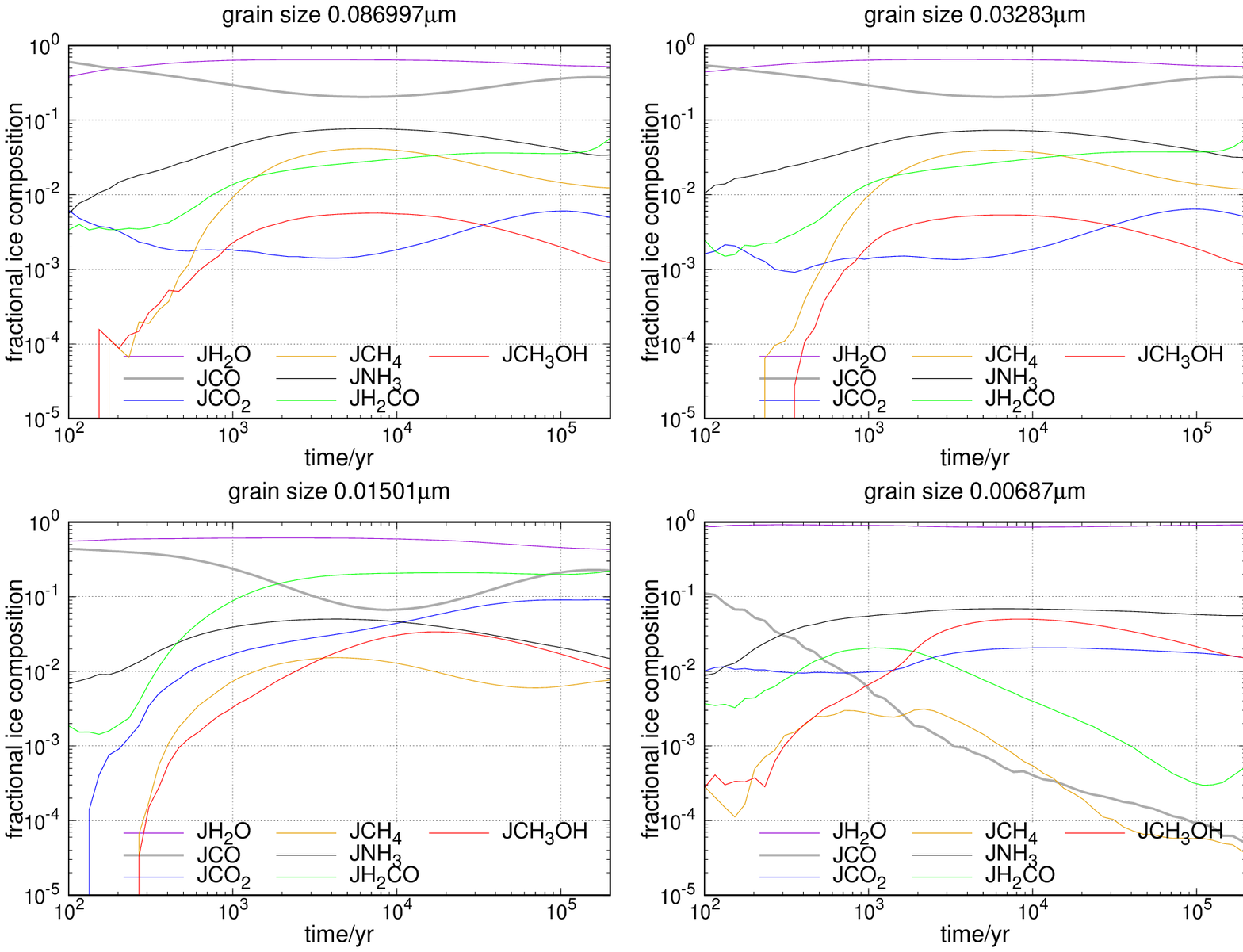}
\caption{The fractional ice composition (showing only major constituents) on different sizes of dust grains as a function of time in model M2.
}
\label{fig:fig_6}
\end{figure*}

Influence of grain heating by cosmic ray induced photons can be clearly seen in Fig.~\ref{fig:fig_6}.
Although high energy photons can heat the second largest grains and the second smallest grains
to more than 12 K and 20 K respectively, the temperature spikes do not alter much the fractional ice composition on those grains.
The fraction of JCO$_2$ on the second smallest grains only slightly increases.
The fractional ice composition on the smallest dust grains are strongly affected by the high energy photons heating.
First, we can see that high energy photons grain heating does not help to produce 
JCO$_2$ on the smallest grains.
The fraction of JCO$_2$ in model M2 on the smallest grains is less than that in model M1. 
Second, we can also see that
as the fractional ice  compositions evolve over time, the fraction of JCO dramatically decreases
in model M2 because JCO can sublimate due to the high temperature spikes induced by
high energy photons in model M2. That also explains why
the extra high energy heating cannot help the production of JCO$_2$ on the smallest dust grains in model M2.
Due to the decrease of the fraction of JCO in model M2, the fraction of JH$_2$CO also decreases in model M2.
The fraction of JCH$_4$ on the smallest dust grains in model M2 also quickly decreases as the fractional ice  
composition evolve over time
because JC atoms can diffuse more quickly to react with surface species other than JH due to the temperature 
spike induced by high energy photons,
thus the number of surface carbon atoms that are hydrogenated to form JCH$_4$ become less on the smallest grains
due to high energy photons heating.

\begin{figure*}
\includegraphics[scale=0.8]{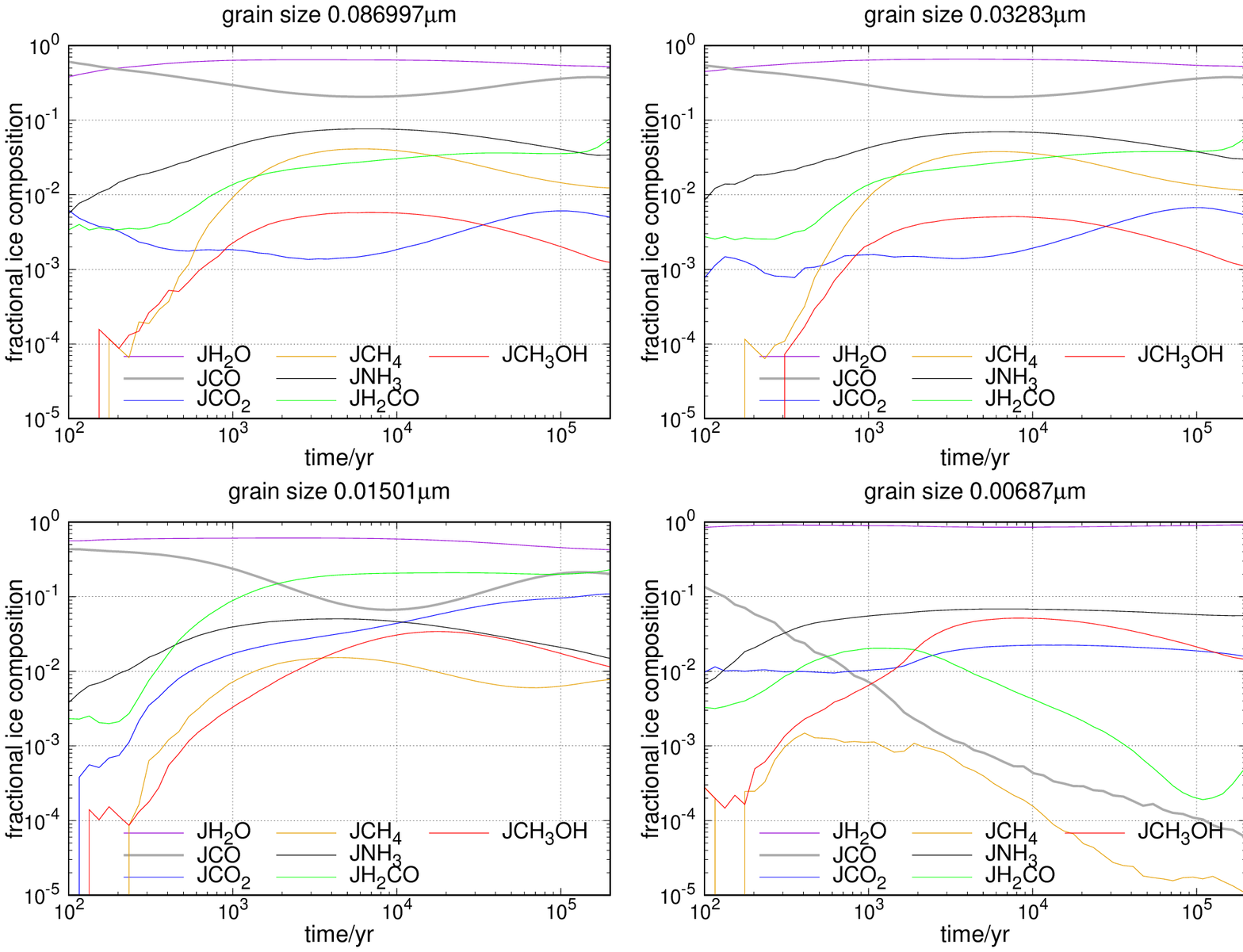}
\caption{The fractional ice composition (showing only major constituents) on different sizes of dust grains as a function of time in model M3.
}
\label{fig:fig_7}
\end{figure*}

Fig.~\ref{fig:fig_7} shows the fractional ice composition (showing only major constituents) 
as a function of time in model M3. By comparing Fig.~\ref{fig:fig_6}
and Fig.~\ref{fig:fig_7}, we can estimate the impact of the approximation that all high energy photons have the same wavelength ( 1300 \AA).
The fractional ice composition as a function of time in models M2 and M3 
are similar. The only difference is that the fraction of JCH$_4$ on the smallest grains in model M2
is around a factor of 4 larger than that in model M3.   
Dust grains in model M3 can be heated to higher temperatures by high energy photons, but the some of 
temperature spikes induced by high energy photons in model M3
are also lower than that in model M2. Thus, models M3 and M2 produce similar results. 

\begin{figure*}
\includegraphics[scale=0.8]{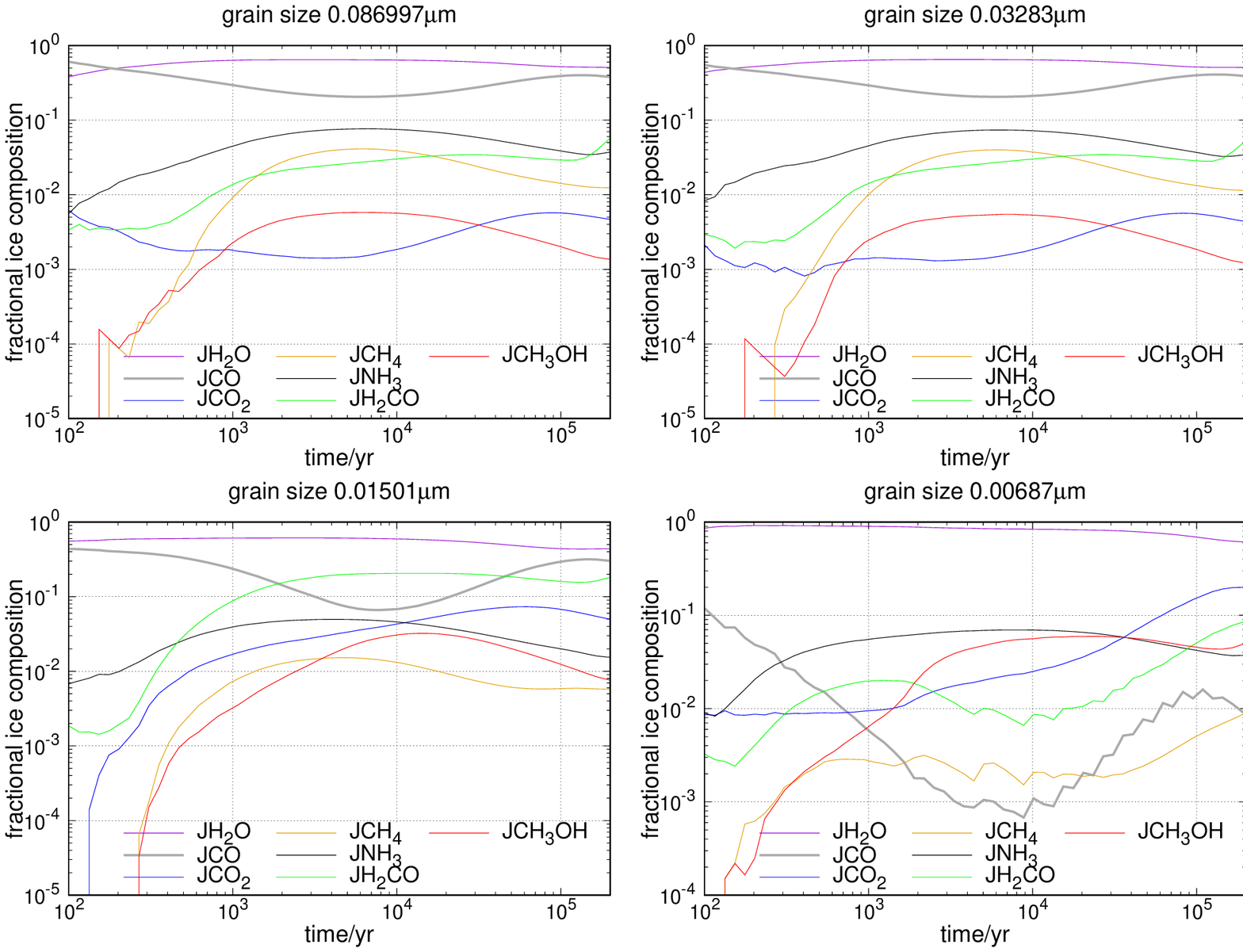}
\caption{The fractional ice composition (showing only major constituents) on different sizes of dust grains as a function of time in model M4.
}
\label{fig:fig_8}
\end{figure*}

The influence of the grain growth on the fractional ice composition (showing only major constituents) can be seen in Fig.~\ref{fig:fig_8}.
The fractional ice compositions on the smallest grains are strongly affected by the grain growth while
the fractional ice compositions on larger grains are not much affected.
Grain growth can decrease the temperature spikes of the overheated smallest grains, thus, offset the effect of high energy photons heating.
We can see that the fraction of JCO gradually increases after $10^4$ yrs, 
Because the radii of the smallest grains become larger, thus the temperature spikes induced by high energy
photon heating become smaller. Similarly,
the fractions of JCO$_2$ and JH$_2$CO increase after $10^4$ yrs of evolution.

\begin{figure*}
\includegraphics[scale=0.8]{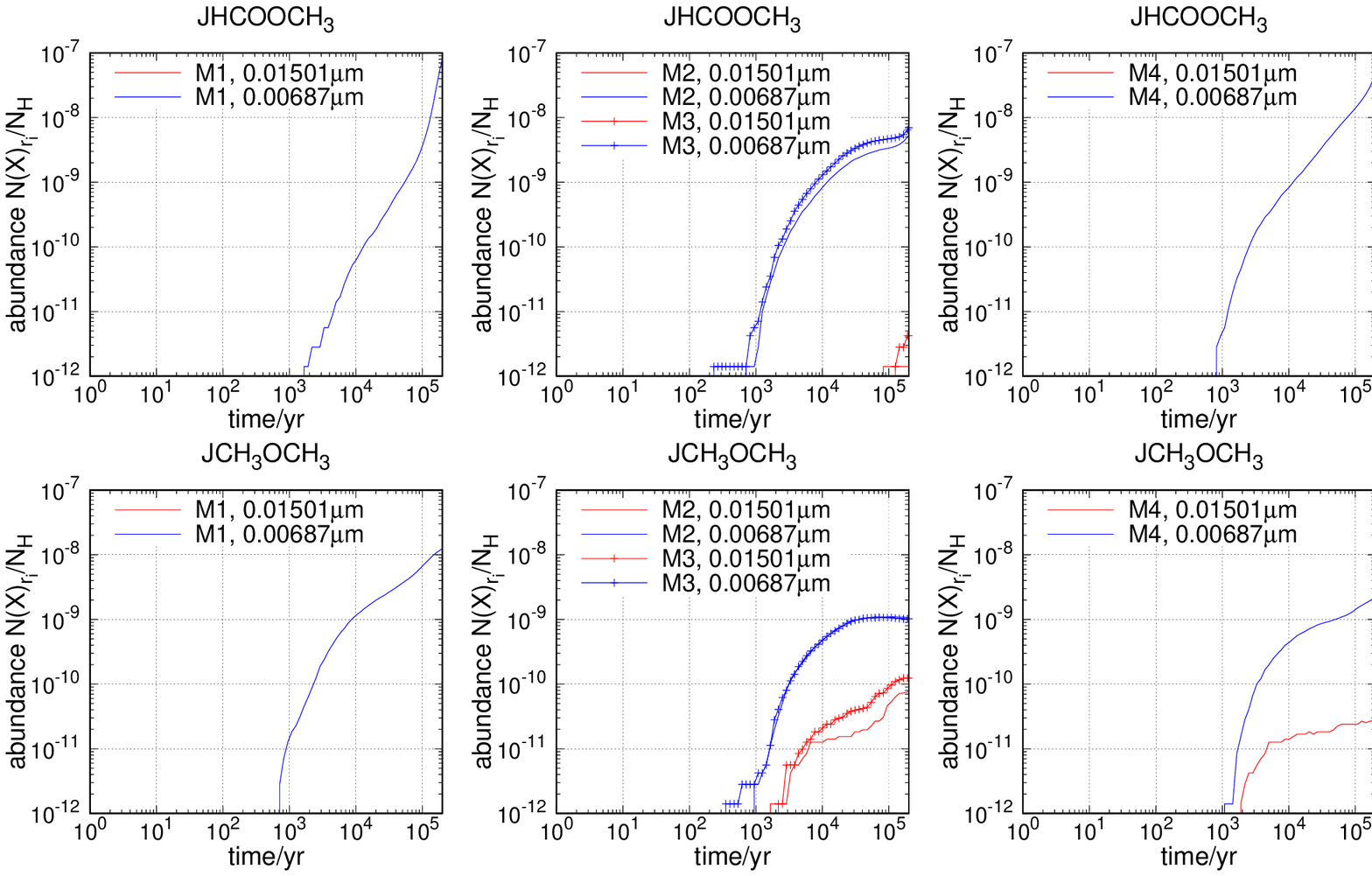}
\caption{The fractional abundance for COMs on different sizes of dust grains as a function of time for different models.
}
\label{fig:fig_9}
\end{figure*}

Small amounts of terrestrial COMs can also be formed on dust grain surfaces in our models from M1 to M4.
Fig.~\ref{fig:fig_9} show the fractional abundances
of JHCOOCH$_3$ and JCH$_3$OCH$_3$ on grains with radius r as a function of time in those models.
In order to form COMs on grain surfaces, the temperature spike of dust grains must be high enough so that radicals can diffuse.
On the other hand, radicals are not easy to be formed on overheated grain surfaces,
so temperature must be low enough so that radicals can be formed on grain surfaces.
In model M1, both JHCOOCH$_3$ and JCH$_3$OCH$_3$ can only be formed on the smallest grains because
radicals that recombine to form JHCOOCH$_3$ and JCH$_3$OCH$_3$ cannot diffuse on larger grains.
The smallest grains are not overheated in model M1 so that sufficient surface radicals which
recombine to form COMs are formed. The effect of high energy photons heating on surface COMs formation is two-fold.
First, in model M2, small amounts of JHCOOCH$_3$ and JCH$_3$OCH$_3$ molecules can also be formed
on the second smallest grains due to the extra high energy photons heating of grains. Second,
the high energy photons heating actually deceases the formation efficiency of JHCOOCH$_3$ and JCH$_3$OCH$_3$ on the
smallest grains because less JCH$_3$O radicals are formed on the smallest grains in model M2.
Model M3 adopts a uniform distribution of the wavelength of high energy photons, however,
COM abundances in model M3 are not much different than that in model M2 in which all high energy 
photons have the same energy.  
As the radii of grains increases, the temperature spikes become less.
So, in model M4, we can see that more JHCOOCH$_3$ and JCH$_3$OCH$_3$
molecules are formed on the smallest grains, however, the formation of JHCOOCH$_3$ and JCH$_3$OCH$_3$ on the
second smallest grains becomes less efficient.

\begin{figure*}
\includegraphics[scale=0.8]{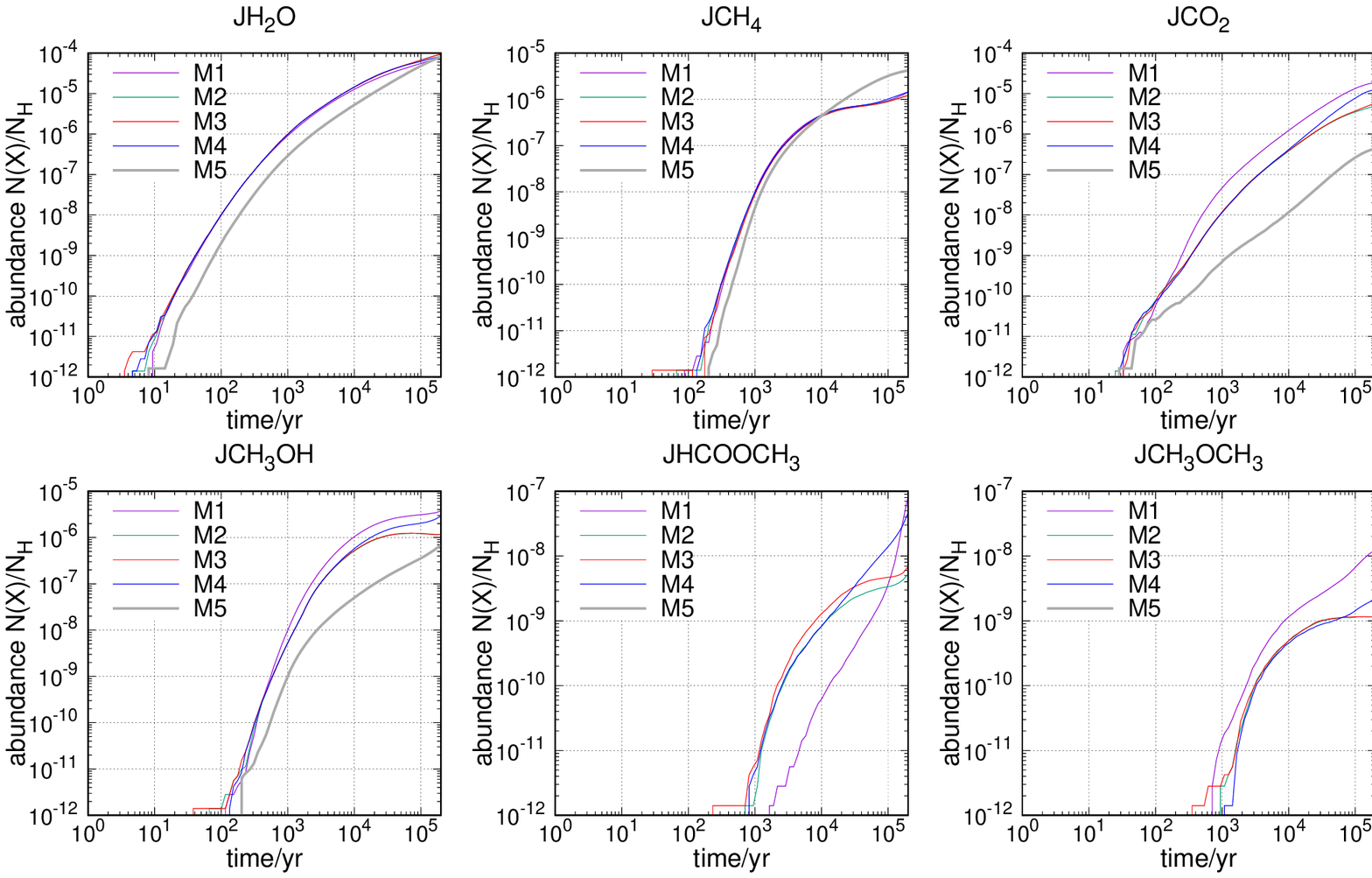}
\caption{The total fractional abundances of major ice mantle species and surface COMs as a function of time.}
\label{fig:fig_10}
\end{figure*}

The total fractional abundances of the major surface species and COMs in all models
as a function of time are shown in Fig.~\ref{fig:fig_10}.
Since grain size distribution is not considered in model M5, the grain surface area in model M5 is less than that in 
other models. Thus, less gas phase species accrete on dust grains in model M5 than that in other models. 
On the other hand, the total fractional abundances of surface species are also strongly affected by the temperature spikes.   
We can see that the total fractional abundances of JH$_2$O and JCH$_4$ are not much affected after a distribution of
grain size and stochastic heating of small grain are included in models. 
However, the total fractional abundances of JCO$_2$ in models M1-4 
are more than one order of magnitude larger than that in model M5 at $2\times 10^5$ yrs.
Moreover, because the smallest grains are overheated in models M2 and M3,
over all JCO$_2$ abundances in models M1 and M4 are higher than that in models M2 and M3.
The total fractional abundances of JCH$_3$OH in models M1-4 are higher than that in model M5
because the temperature spikes help to form methanol as explained before and the smaller grain surface area in model M5.
Moreover, the total fractional abundances of JCH$_3$OH in models M2 and M3 is lower than that in models M1 and M4
because the smallest grains are overheated in models M2 and M3 so that the formation of surface methanol in models M2 and M3 is
less efficient than that in models M1 and M4.   
The terrestrial COMs, JHCOOCH$_3$ and JCH$_3$OCH$_3$, can be formed in models from M1 to M4, but not in M5 in which
the dust grain temperatures are kept to be 10 K because its formation requires the surface diffusion of radicals. 

\section{Comparison with observations and other models}\label{sec:compare}
In this section, we compare our model results with the observational ice compositions towards background star Elias 16
and the model results from previous studies. The selected models are MRN2 model in \citet{Acharyya2011} and
5G\_T10\_DIST in \citet{Pauly2016} because these two models adopt the same grain size distribution 
as that used in this work. 

The comparison is shown in the Tab.~\ref{tab:obs}. Water ice abundances in all models are in good agreement with the observed values.
The MRN2 model only includes a grain size distribution and grain growth while 
the temperatures of grains are fixed to be 10 K.
Therefore, JCO$_2$ abundances in the model MRN2 are more than one order of magnitude less than the observed values.
The model 5G\_T10\_DIST includes the temperature distribution of dust grains so that the temperatures of 
small dust grains are more than 12 K, so JCO$_2$ abundances in the 5G\_T10\_DIST is slightly more than the observed abundance.    
Our models M1, M2 and M3 take into account of the temperature fluctuations so that the temperature 
of small dust grains may exceed 12 K, therefore
significant amounts of JCO$_2$ can also be produced. The abundances of JCO$_2$ in models M1 and M4 are close to the observed
value while JCO$_2$ abundances in models M2 and M3 are about a factor of 4 smaller than the observed abundance.
We can also see that abundances of JCO in our models  M1-4 
are larger than any other previous models and are slightly more abundant than the observed value.
The observed abundances of JNH$_3$ and JCH$_3$OH towards background star Elias 16 have upper limits only 
while most our model results are below this upper limit except that JCH$_3$OH in models M1 and M4
is slightly more abundant than the observed value. The abundances of JCH$_3$OH in our models
is about the same as that in the model 5G\_T10\_DIST, but less than that in model MRN2. 
Our models from M1 to M4 produce less JNH$_3$ than previous models do.
Surface JCH$_4$ and JH$_2$CO have not been detected toward Elias 16.
We can see that the abundance of JCH$_4$ in the models M2 and M3 is the lowest among all models
while the abundances of JH$_2$CO in all our models from M1 to M4 are more than one order of magnitude higher 
than previous model results.

\begin{table*}
\caption{Comparison with observed ice compositions and previous model results.}
\label{tab:obs}
\begin{tabular}{lllllllll}
\hline
species    &  Elias 16   &  MRN2      & MRN2    & 5G\_T10\_DIST   & M1              & M2               &  M3             & M4       \\
time (yrs) &             &  $10^5$    & $10^6$  & $5\times10^6$   & $2\times10^5$   & $2\times10^5$    & $2\times10^5$  & $2\times10^5$  \\  
\hline
JH$_2$O    &  6.4(-5)    &  1.85(-5)  & 7.0(-5) &  1.59(-4)       & 7.72(-5)        &  9.43(-5)   &  9.40(-5) & 8.24(-5) \\
JCO        &  26         &  29        & 2(-3)   &  7.9            & 34              &  39         & 38       &  47       \\
JCO$_2$    &  20         &  1.4       & 0.27    &  36.2           & 24              &  5          & 6        & 15       \\
JCH$_4$    &  -          &  14        & 28      &  23.6           & 2               &  1          &  1        &  2        \\
JCH$_3$OH  &  $<3$       &  4.4       & 16      &  2.9            & 5               &  1          & 1        &  4        \\
JH$_2$CO   &  -          &  1.5(-4)   & 8.2(-8) &  0.42           & 23              &  15         & 15       &  20       \\
JNH$_3$    &  $<9$       &  11        & 8.1     &  20.5           & 6               &  6          & 6        & 6        \\
\hline
\end{tabular}
\medskip{\protect\\
Notes.\protect\\
a(-b) means $a\times10^{-b}$. Observational Data are collated by \citet{Garrod2007}. 
JH$_2$O abundances are respect to H nuclei abundances while the 
abundances of other species are given in the percentage of water ice abundances.
The original JH$_2$O abundances in model MRN2 were respect to H$_2$ and we have converted the abundances
to that respect to H nuclei.
\protect\\
}
\end{table*}

Small amounts of COMs can be formed on the smaller grains in our models M1-4.
The total fractional abundance of JHCOOCH$_3$ and JCH$_3$OCH$_3$ can be as high as a few $10^{-8}$ and $10^{-9}$
respectively in model M4, which includes both stochastic heating of dust grains and grain growth. 
The formation of COMs on grains was not discussed in the previous models \citep{Acharyya2011,Pauly2016}.
However, we do not expect COMs can be formed in previous models because the temperatures of even the smallest grains are below 20 K,
so radicals can hardly diffuse to form COMs.

\section{Discussion and Summary}\label{sec:sum}
We use the macroscopic Monte Carlo method to simulate gas-grain reaction networks under physical conditions pertain to
cold dark clouds. A distribution of grain sizes and stochastic heating of the smaller dust grains are included in the simulations. 
Five models are studied in this work. We simulate three models M1, M2 and M3 that do not include grain growth.
Model M1 only considers dust grain heating by ambient low energy photons while M2 and M3 also include the dust grain heating
by cosmic ray induced UV photons inside molecular cloud.
All cosmic ray induced UV photons have the same energy in model M2 while the wavelength of cosmic ray induced UV photons follows
a uniform distribution in model M3.
Model M4 is simulated in order to find the effect of grain growth on the chemical evolution of ice mantles on stochastically heated grains.
Finally, a reference model M5, in which there is no grain size distribution or stochastic heating of grains, 
is simulated for comparison.

The fluctuations of the grain temperature dramatically alter the ice mantle compositions 
on the smaller dust grains. The abundances of JCO, JCO$_2$, and COMs are more dependent on the temperature 
fluctuation of dust grains than any other species.
The abundances of JCO$_2$ increase significantly when stochastic heating is included in the models, 
which agree with previous studies which adopted size-dependent grain temperatures~\citep{Pauly2016}. 
Moreover, our simulation results show that models that include low energy photons only can already produce
enough JCO$_2$ that is consistent with the observational data toward Elias 16. Small amounts of terrestrial COMs can 
also be produced on the smallest grains due to the temperature spikes induced by low energy photons.  
The influence of high energy photons heating in models is two-fold. First, the smallest grains are overheated
so that less JCO$_2$ or COMs are formed. Moreover, JCO can easily sublime on the smallest grains.
Second, the temperature spikes on the second largest grains are high enough so that more JCO$_2$ and COMs
are formed.
The complicated spectrum of high energy photons is approximated as either a single photon energy or a uniform distribution in our models.
The choice of approximation does not have a large impact on the simulation results.
Grain growth, which is included in model M4, is able to decrease the temperature spikes of the overheated dust grains as ice mantles 
gradually accumulate on dust grains. Thus, more COMs and JCO$_2$ are formed on the smallest grains 
which are overheated by high energy photons.

\citet{Schutte1991} argued that JCO sublimation cooling may be more important than 
radiative cooling if the temperatures of dust grains are above 26K. Their conclusion is based the assumption
that the whole grain surface is covered by volatile species (JCO). 
The overheated grains should cool down much more quickly if JCO sublimation cooling dominates over radiative cooling.
However, we argue that JCO sublimation cooling is not likely to be important in our models because of the following reasons.
The fraction of JCO on the smallest grains is less than $10^{-3}$ at the time $10^4$ yrs 
in model M2 as shown in Fig.~\ref{fig:fig_6}. The total population of surface species on the smallest grains 
at the time $10^4$ yrs is a few monolayers. Therefore, the fraction of the surface covered by JCO on the smallest grains
at the time $10^4$ yrs is less than 1\% in model M2. Similarly, we can estimate that the fraction is also less than  1\%
at other time steps. Since the number of JCO desorbed from grain surfaces within fixed time interval is 
proportional to the population of JCO on grain surfaces,
following \citet{Schutte1991}, we can estimate that the energy taken away by JCO sublimation on the smallest grains is less that 1\% of 
that by radiative cooling at 26 K. So radiative cooling still dominates over JCO sublimation cooling at 26 K because of the low coverage of JCO.   
Moreover, we can also estimate how much JCO sublimation cooling can decrease the temperature spike of the smallest bare grains in model M2.
Following \citet{Schutte1991}, we assume the energy taken away by the sublimation of each JCO molecule is 960 K. 
Solving Equ.~\ref{equ0}, the temperature of the smallest bare dust grain drops by $\Delta T\sim 0.24$ K for each JCO sublimation event at T = 26 K.
Equ.~\ref{equ0} shows that the heat capacity of grains increases as T increases, thus $\Delta T $ should decrease at higher temperatures.
On the other hand, we can estimate that each high energy photon can only desorb a few 
JCO molecules from each of the smallest grains.
We found that less than 80 thousand JCO molecules desorb from each of the smallest dust grains during $2\times 10^5$ yrs in model M2
while each of the smallest grains are bombarded by around 30 thousand high energy photons during the same time period.
Since JCO can hardly sublime due to the temperature spike induced by low energy photons,
we can approximately estimate that each high energy photon can only desorb less than three 
JCO molecules from each of the smallest grains.
So JCO sublimation cooling can only lead to a temperature drop of $<$ 0.72 K.
Therefore, we can conclude that JCO sublimation cooling may not have a large impact in model M2. 
Moreover, the JCO sublimation cooling should be even less significant in models 
that include grain growth because less JCO molecules sublime in these models. 

It is particularly interesting that COMs can be formed on the smallest grains when stochastic heating is included in models.
The amount of COMs formed in models is small however. The fractional abundances of JHCOOCH$_3$ and JCH$_3$OCH$_3$
are a few $10^{-8}$ and $10^{-9}$ respectively. The reason is that the abundances of radicals which can recombine 
are small on the smallest grains. Recent study shows that radicals can accumulate in ice mantles if we adopt a 
three phase model with photon penetration and bulk diffusion~\citep{Chang2014}. Further research is necessary to 
study how COMs can be formed inside ice mantle via bulk diffusion mechanism. 

Traditionally, dust grains are assumed to be uniform in size with radii $0.1\mu m$, so
the temperature fluctuations of dust grains are ignored in most astrochemical models. 
Our simulation results show that the temperature fluctuations of small grains 
whose radii are about $0.006\mu m$ are large enough to alter the compositions of ice mantle. 
On the other hand, because of dust coagulation, the population of grains which are small enough  
to undergo significant temperature fluctuations may be less than that predicted by the MRN grain size distribution.
Therefore, more study should be done to investigate the roles of small grain for the evolution of molecular clouds.

We summarize our main results as the following:
\begin{enumerate}[1.]
\item The temperatures of small grains inside cold molecular clouds are strongly affected by the ambient interstellar radiation
and cosmic ray induced secondary photons. Considering the ambient interstellar radiation only,
the temperature spikes are less than 16 K for the grains with radius $0.01501\mu m$ while
the temperature spikes can be more than 20 K for the smaller grains with radius $0.00687\mu m$.
The cosmic ray induced secondary photons can further increase the grain temperature fluctuation.
The highest temperatures for grain with radii $0.01501\mu m$ and $0.00687\mu m$ can be more than 20 K and 35 K respectively.
\item Dust grain temperature fluctuations can increase the production of JCO$_2$. 
Over all, the formation efficiency of JCO$_2$ on grain surfaces decreases if the 
cosmic ray induced secondary photons heating is included in models. 
\item The abundances of JCO on the smallest grains that are heated by cosmic ray induced secondary photons are much smaller
than that on other sizes of grains because JCO can sublimate on these grains.
\item Small amounts of terrestrial COMs are able to be formed on stochastically heated small grains because radicals that
recombine to form COMs are mobile because of the temperature spike. 
\item Grain growth is able to decrease the temperature spikes of the overheated smallest dust grains, thus, the formation of JCO$_2$ and COMs on the smallest grains
becomes more efficient as the size of grains increases in models that include high energy photons heating. 
\item The grain mantle chemistry is not much affected if a model adopts a uniform distribution of the wavelength of high energy photons 
instead of the median wavelength.  

\end{enumerate}

\section*{Acknowledgments}
Q. Chang is a research fellow of the One-Hundred-Talent project of the Chinese Academy of Sciences.
This work was funded by The National Natural Science foundation of China under grant 11673054.
We thank Eric Herbst for critical reading of the manuscript and helpful discussions.
The Taurus High Performance Computing system of Xinjiang Astronomical Observatory was used for the simulations.
We thank our referee's very constructive comments to improve the quality of the manuscript.


\bsp

\label{lastpage}

\end{document}